\documentclass{aa}
\usepackage{graphics}

\begin{document}


   \title{The stability of the Circumnuclear Disk clouds\\ in the
          Galactic Centre}

   \titlerunning{The stability of the CND clouds in the GC}

   \author{B.~Vollmer\inst{1,2} \and  W.J.~Duschl\inst{2,1}}

   \offprints{B.~Vollmer, e-mail: bvollmer@mpifr-bonn.mpg.de}

   \institute{Max-Planck-Institut f\"ur Radioastronomie, Auf dem
              H\"ugel 69, 53121 Bonn, Germany. \and
              Institut f\"ur Theoretische Astrophysik der
              Universit\"at Heidelberg, Tiergartenstra{\ss}e 15,
              69121 Heidelberg, Germany.}

   \date{Received / Accepted}

\abstract{The influence of rotation and magnetic fields
on the physical properties of isothermal gas clouds is discussed. 
The presence of rotation and/or magnetic fields
results in an increase of the critical cloud mass with respect
to gravitational instability for clouds of a
given temperature and external pressure. Rotating clouds have
higher densities. Consequently, they are more stable against tidal
shear than non-rotating clouds. They can approach the Galactic
Centre up to a radius of $\sim$2~pc without being disrupted by the
tidal shear due to the gravitational potential. For smaller radii
the clouds either collapse or become tidally disrupted. We suggest
that this mechanism is responsible for the formation of the inner
edge of the Circumnuclear Disk in the Galactic Centre.  }

\maketitle

\section{Introduction}

The Galactic Centre\footnote{We assume 8.5 kpc for the distance to
the Galactic Centre.} is surrounded by a
large number of gas and dust clouds. The distribution
and the kinematics of these clouds are generally interpreted
as a thick disk ({\it Circumnuclear Disk CND}). Whether this disk--like
structure has an outer edge is still a matter of debate. The CND
was studied in molecular lines by Gatley et al. (1986)
(H$_{2}$), Serabyn et al. (1986) (CO, CS), G\"{u}sten et al.
(1987) (HCN), DePoy et al. (1989) (H$_{2}$), Sutton et al. (1990)
(CO), Jackson et al. (1993) (HCN), and Marr, Wright, \&
Backer (1993) (HCN), Coil \& Ho (1999, 2000) (NH$_{3}$),
Wright et al. (2001) (HCN). The disk has a sharply defined inner 
edge at a radius of $\sim$1.7~pc. Inside this radius the gas density 
drops by more than one order of magnitude. G\"{u}sten et al. (1987)
suggested that the density of the clouds is not high enough to
resist tidal shear due to the centre's strong gravitational field.
Therefore, they concluded that the CND is a very short lived
structure ($\sim 10^{5}$~yr). On the other hand, Jackson et al.
(1993) pointed out that the possibility should not be
dismissed that the clouds' density is even high enough to
stabilize individual gas clouds against tidal disruption ($n\sim
10^{7}$~cm$^{-3}$). In a previous paper (Vollmer \& Duschl
2001; hereafter VD2001) we constructed an analytic model for a
clumped gas and dust disk and applied it to the CND. The clouds
were treated as isothermal selfgravitating spheres with a 
given outer pressure. We
succeeded in reproducing the main characteristics of the observed
gas clouds (mass, central density, density at the outer boundary).
A major shortcoming of this model was that these clouds could not
resist tidal shear at distances smaller than $\sim$2.5~pc from the
Galactic Centre. In this work we investigate the influence of
rotation on the cloud properties and their stability against tidal
shear.

\section{Basic Picture}

As in VD2001 we assume that during a short accretion event 
($\Delta t \sim 10^{6}$ yr) an amount of gas of several 10$^{4}$ M$_{\odot}$ 
is driven into the Galactic Centre region to distances less than 10~pc. 
The infalling gas has not a uniform density distribution and might be 
turbulent. This clumpy medium is exposed to the ambient
UV radiation field due to the population of young O/B stars in the Galactic
Centre. Low density regions of the infalling gas are evaporated 
rapidly while regions of higher density stay molecular and are heated 
to an equilibrium temperature during less than an orbital period.
In this way a clumpy circumnuclear disk is formed. We assume that
the clouds are gravitationally stable. Three kinds of pressure can
counterbalance the selfgravitation of the clouds:

(i) Turbulent pressure:
if the turbulent pressure is dominant, we can estimate the turbulent
velocity dispersion $\sigma$ with the help of the Virial theorem:

\begin{equation}
\sigma^{2} \sim \frac{2GM_{\rm cl}}{r_{\rm cl}}\ ,
\end{equation}
where $M_{\rm cl}$ is the cloud mass and $r_{\rm cl}$ its radius.
With $M_{\rm cl}$=30~M$_{\odot}$ and $r_{\rm cl}$=0.05~pc, the
velocity dispersion is $\sigma \sim 2.3$~km\,s$^{-1}$ 
(note that the observed FWHM is two times $\sigma$).

(ii) Magnetic pressure:
in the case of dominant magnetic pressure, we can estimate
an equivalent linewidth $c$ using a local magnetic field strength
$B$=3~mG (Yusef-Zadeh et al. 1996).

\begin{equation}
B^{2}/8\pi \simeq c^{2}\rho_{\rm cl}\ ,
\end{equation}
where $\rho_{\rm cl}$ is the cloud density. With a cloud
density of $\rho_{\rm cl}$=3\,10$^{-18}$~g\,cm$^{-3}$ we obtain 
an equivalent linewidth of $c \sim 3.3$~km\,s$^{-1}$.

(iii) Thermal pressure:
if the thermal pressure dominates, the sound velocity at
a temperature of 150~K is $c_{\rm s} \sim 0.8$~km\,s$^{-1}$.
Thus, the linewidth $c$ of a single selfgravitating cloud
is 1~km\,s$^{-1} \leq c \leq 3$~km\,s$^{-1}$.

The observed linewidths are as large as $\Delta v = 50$~km\,s$^{-1}$
(see, e.g., Wright et al. 2001). If the clouds within the CND
are gravitationally stable, their intrinsic linewidth is much
smaller. A possible explanation of the large observed linewidths  
is an enhanced turbulent velocity dispersion between the clouds due
to infalling gas streamers and/or superposition of clouds
with different rotational and turbulent velocities
(the turbulent velocity dispersion of the model disk
of VD2001 is of the order $\Delta v \sim 15$~km\,s$^{-1}$).
The study of a CS(3--2) data cube observed with the IRAM 30m
Telescope (Zylka et al. 1999)
indicates that the features of linewidths up to 50~km\,s$^{-1}$ might
consist of several clouds with small individual
linewidths ($\sim 2$~km\,s$^{-1}$) (Zylka, private communication).

In the following we discuss rotating clouds without a magnetic field
first. As shown above, the inclusion of a magnetic field increases
the equivalent linewidth of the clouds by a factor $\sim$3. 
The implications of the inclusion of a magnetic field
will be discussed in Sect.~\ref{sec:magneticfield}.

\section{Cloud Rotation}

The equilibrium state of rotating isothermal clouds was studied
both analytically (Hayashi, Narita, \& Miyama 1982; Tohline 1985a,
b) and numerically (Stahler 1983a, b; Kiguchi et al. 1987; Narita
et al. 1990). These authors showed that the density and the mass
of rotating clouds can exceed that of non-rotating clouds
considerably. Kiguchi et al. (1987) carried out numerical
simulations of rotating isothermal spheres embedded in a tenuous
intercloud medium, i.e. the outer boundary of the cloud is determined
by the external pressure. Their calculations did not include 
a magnetic field. They concluded that a rotating cloud is
dynamically stable if

(i) $M_{\rm crit}^{\rm rot}/M_{\rm crit}^{\rm BE} < 31$, where
$M_{\rm crit}^{\rm rot}$ is the critical mass for gravitational
instability of the rotating cloud and $M_{\rm
crit}^{\rm BE}$ is the critical mass for gravitational
instability of a non-rotating
Bonner--Ebert sphere with the same sound speed (temperature) and
the same outer pressure;

(ii) the maximum mean rotation velocity is smaller than
2.7\,$c_{\rm s}$, where $c_{\rm s}$ is the sound velocity;

(iii) $\overline{\rho}/\rho_{\rm ext} < 6$, where
$\overline{\rho}$ is the mean density of the cloud and $\rho_{\rm
ext}$ is the external density.

Thus, rotating isothermal clouds, which are gravitationally
stable are denser and more massive than
non-rotating clouds of the same temperature embedded in a medium
of the same pressure. We conclude that cloud rotation alters
the solution for the physical characteristics of our model clouds
(VD2001) in such a way that more massive and denser clouds
result which are more stable against tidal disruption.

\section{Specific angular momentum}

\subsection{Cloud formation \label{sec:cloudformation}}

The spin angular momentum of a cloud in the CND should reflect the
angular momentum of the interstellar medium from which the cloud
has formed (see e.g. Blitz 1993). Within the central
10~pc the gravitational potential is approximatly spherical.
The rotation curve is determined for distances smaller than
approximatelly 1~pc by the gravitational potential of the central black 
hole and for distances greater than this by that of the nuclear star 
cluster (Genzel et al. 1996).
If we consider that a cloud
forms when a disk-like region becomes gravitationally unstable and
collapses, we expect a similar result to that considered
analytically by Mestel (1966). Since the rotation curve in the
Galactic Centre is approximately constant for distances greater 
1~pc and is falling inwards for smaller distances, the rotation of the
cloud can be either prograde or retrograde depending on the
details of the cloud formation. We thus have to estimate the
specific angular momentum of the ISM at the place of cloud
formation. For an isolated CND, VD2001 found a very small, 
mean, radial drift velocity of the clouds 
($v_{\rm drift}< 0.01$~km\,s$^{-1}$; however,
a single cloud can approach the Galactic Centre faster). 
In this case we can assume
that the clouds were formed not far away from the galactic
distances where we observe them today ($2\,{\rm pc} \leq {\rm
R}_{\rm G} \leq 7$\,pc). Mestel (1966) has shown that a cloud can
have a specific angular momentum about its mass-centre up to
$J/M_{\rm cl}=0.5R^{2}\Omega$, where $R$ is half the size of the
collapsed region and $\Omega$ is the local angular velocity. Let
us consider a cloud which forms at a distance greater than 2\,pc
from the Galactic Centre. With a typical density in the disk
$\rho=2\,10^{-19}$\,g\,cm$^{-1}$ and a typical cloud mass of
30\,M$_{\odot}$ (VD2001), the size of the collapsing region is
$R\simeq$0.1\,pc. The initial value of $\Omega$ depends on the
details of the collapse. With $\Omega(2\,{\rm pc})\leq$
60\,km~s$^{-1}$~pc$^{-1}$ we obtain a specific spin angular
momentum of $J/M \leq$ 0.3 pc\,km\,s$^{-1}$. 
If the CND is not isolated, the clouds could have 
formed at higher distances from the Galactic Centre ($\sim$10~pc).
Using $\Omega(10\,{\rm pc})\sim$12\,km~s$^{-1}$~pc$^{-1}$,
gives $J/M \leq$ 6\,10$^{-2}$ pc\,km\,s$^{-1}$. In addition,
partially inelastic off-center collisions between the clouds can
lead to changes in the spin angular momentum of the clouds.

\subsection{Stability criteria}

Limits on the specific angular momentum of the clouds $J/M$ are
given by the onset of bar formation on the one hand and the onset
of core collapse on the other hand. Kiguchi et al. (1987)
suggested that bar formation occurs for $\beta_{0}> 1/3$, where

\begin{equation}
\beta_{0}=\frac{25}{12}\big(\frac{4\pi}{3}\big)^{\frac{1}{3}}
\frac{\rho_{\rm ext}^{\frac{1}{3}}J^{2}}{GM^{\frac{10}{3}}}
\end{equation}
is the ration of rotational energy to gravitational energy ratio of the
cloud. Here $\rho_{\rm ext}$ is the boundary density of the cloud,
$J$ is the angular momentum of the cloud, $M$ its mass, and $G$
the gravitation constant. The criterion for clouds which are
stable against bar formation is thus

\begin{equation}
J/M \leq 0.31 G^{\frac{1}{2}} \rho_{\rm ext}^{-\frac{1}{6}}
M^{\frac{2}{3}} \sim 7\,10^{-2}\ {\rm pc\,km\,s}^{-1}\ ,
\end{equation}
where we have used typical cloud parameters ($M$=30~M$_{\odot}$,
$\rho_{\rm ext}=3\,10^{-19}$~g\,cm$^{-3}$). 
Miyama, Hayashi, \& Narita (1984) found that
rotating isothermal spheres collapse if $\alpha_{0}\beta_{0} <
0.2$, where

\begin{equation}
\alpha_{0}=\frac{5}{2}\big(\frac{4\pi}{3}\big)^{-\frac{1}{3}}
\frac{c_{\rm s}^{2}}{G\rho_{\rm ext}^{\frac{1}{3}}M^{\frac{2}{3}}}
\end{equation}
is the thermal to gravitational energy of the cloud. The criterion
for stable clouds thus is

\begin{equation}
J/M \geq 0.2 \frac{GM}{c_{\rm s}} \sim 5\,10^{-2}\ {\rm
pc\,km\,s}^{-1}\ ,
\end{equation}
using the cloud parameters from above and $c_{\rm s}$=1~km\,s$^{-1}$.

In Sect.~\ref{sec:magneticfield} it is shown that the presence of
magnetic fields with a field strength of several mG results in
an effective linewidth of $c \sim 3$~km\,s$^{-1}$. In this case
the specific angular momentum is 
\begin{equation}
J/M \geq 0.2 \frac{GM}{c} \sim 2\,10^{-2}\ {\rm
pc\,km\,s}^{-1}\ ,
\end{equation}
We thus conclude that for a stable rotating cloud $2\,10^{-2} \leq
J/M \leq 7\,10^{-2}$~pc\,km\,s$^{-1}$. This is consistent with 
the upper limits $6\,10^{-2} \leq J/M \leq 0.3$~pc\,km\,s$^{-1}$ at 
the moment of cloud formation (Sect.~\ref{sec:cloudformation}).

\section{A typical rotating cloud}

The maximum central density $\rho_{\rm c}$ of an isothermal
cloud with external pressure $P_{\rm ext}$ (Bonner--Ebert
sphere) is $\rho_{\rm c}/\rho_{*}= 14$, where
$\rho_{*}=c_{\rm s}^{-2}P_{\rm ext}$. Stable rotating
isothermal clouds can have central densities up to $\rho_{\rm
c}/\rho_{*}\sim 100$. Fig. 14 of Kiguchi et al. (1987) shows the
region of stable clouds as a function of the central density and
the cloud mass. In our case 
$M_{*}=c_{\rm s}^{4}P_{\rm ext}^{-\frac{1}{2}}G^{-\frac{3}{2}}$=15~M$_{\odot}$
and $\rho_{*}=10^{-18}$~g\,cm$^{-3}$. Since we are
interested in rotating clouds that have higher central densities
than non-rotating clouds, we will only discuss the region for
$\rho_{\rm c}/\rho_{*}>14$. In this case our region of interest is
limited by the dashed collapse curve and the dash-dotted curve for
bar formation in Fig.~14 of Kiguchi et al. (1987). A typical cloud
in this region has $M/M_{*} \sim 2-3$ or $\big(\rho_{\rm
c}/\rho_{*}\big)^{\rm rot}\sim 3\,\big(\rho_{\rm
c}/\rho_{*}\big)^{\rm BE}$. Such a cloud has the following
physical characteristics (Kiguchi et al. 1987; Table 3):
$\beta_{0}=0.16,\ \rho_{\rm c}/\rho_{*}=30,\ M/M_{*}=2.38,\
c_{\rm s}J/(GM^{2})=0.164,\ \overline{\rho}/\rho_{*}=2.71$.
In physical units this gives: $M$=36~M$_{\odot}$,
$\overline{\rho}=2.7\,10^{-18}$~g\,cm$^{-3}$, 
$\rho_{\rm c}=3\,10^{-17}$~g\,cm$^{-3}$. The radius perpendicular to the
rotation axis is $r_{\rm e}=8\,10^{-2}$~pc and parallel to the
rotation axis $z=3\,10^{-2}$~pc. The maximum mean rotation
velocity is $v_{\rm rot}\sim 0.8$~km\,s$^{-1}$.

This cloud has thus a 3 times higher central density than a
non-rotating isothermal selfgravitating sphere of the same temperature 
which is embedded in a tenuous medium of the same outer pressure.

The clouds are illuminated by the UV radiation field coming from
the central He\,{\sc i} star cluster. The mass loss rate of
these clouds due to outflowing ionized gas is approximately
$\dot{M} \sim 4\pi r_{\rm cl}^{2} \rho_{\rm cl} c_{\rm i}$, where
$r_{\rm cl}$ is the cloud radius and $c_{\rm i}$ is the sound
velocity in the external ionized gas. Since for the diameter of
such a cloud $d_{\rm cl}^{\rm rot} < 2\times d_{\rm cl}^{\rm BE}$,
rotation lowers the lifetime of a cloud by a factor smaller than
4. The minimum lifetime is still greater than 10$^{7}$~yr.

\section{Magnetic fields \label{sec:magneticfield}}

Tomisaka et al. (1989) studied the structure of selfgravitationally
stable magnetized clouds. They showed that the critical mass
for stability is 

\begin{equation}
M_{\rm mag} \sim 1.18\Big(1-\big(\frac{0.17}{G^{\frac{1}{2}}|{\rm d}M_{\rm cl}
/{\rm d}\Phi_{\rm B}|_{\rm c}}\big)^{2}\Big)^{-\frac{3}{2}}\,M_{\rm BE}\ ,
\end{equation}
where $|{\rm d}M_{\rm cl}/{\rm d}\Phi_{\rm B}|_{\rm c}$ represents the 
mass-to-magnetic flux ratio at the center and $M_{\rm BE}$ is the
Bonner--Ebert critical mass in the absence of magnetic fields.
If we approximate $|{\rm d}M_{\rm cl}/{\rm d}\Phi_{\rm B}|_{\rm c}
\simeq M_{\rm cl}\big(B r_{\rm cl}^{2}/2\big)^{-1}$ and let 
$M_{\rm cl}=$30~M$_{\odot}$,
$r_{\rm cl}$=0.05~pc, and $B$=5~mG, we obtain $M_{\rm mag} \sim
2.5 M_{\rm BE}$. This is consistent with a realistic Bonner--Ebert mass of 
$M_{\rm BE} \sim 10$~M$_{\odot}$ (VD2001).
We would thus expect that the density of a magnetized cloud of size
$d_{\rm cl}$=0.1~pc, mass $M_{\rm cl}$=30~M$_{\odot}$, and magnetic
field strength $B$ of several mG would also show a density
enhancement of a factor 3 with respect to a non-rotating, non-magnetized
isothermal, selfgravitating cloud.

\section{Tidal shear}

\begin{figure}
    \resizebox{\hsize}{!}{\includegraphics{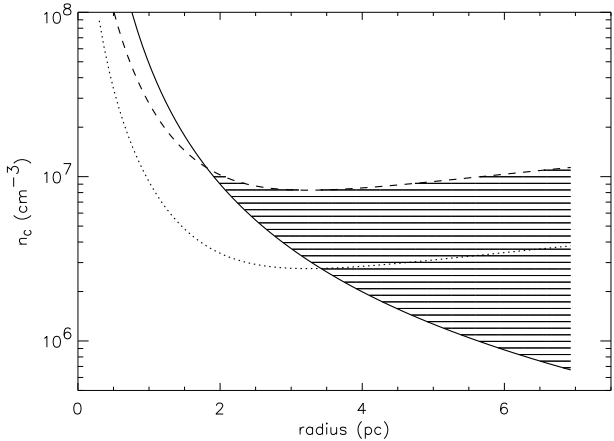}}
      \caption{ \label{fig:model_tidal}
    Central density of the clouds as a function of the distance
    from the Galactic Centre. Solid: critical density for tidal
    disruption. Dotted: model central density of VD2001.
    Dashed line:  central density of a rotating cloud.
    Dashed surface: range of densities
    where clouds are gravitationally and tidally stable.
    }
\end{figure}

The critical density for a cloud orbiting at a distance $R$ on a
circular orbit around a point mass $M$ below which it will
disintegrate due to tidal shear is given by
\begin{equation}
\rho_{\rm crit}=\frac{3}{2\pi}\frac{M}{R^{3}}
\end{equation}
(see e.g. Stark et al. 1989). We approximate the mass distribution
in the Galactic Centre by $M(R)=M_{0}+M_{1}\,R^{1.25}$, where
$M_{0}=3\,10^{6}$~M$_{\odot}$ an
$M_{1}=1.6\,10^{6}$~M$_{\odot}$\,pc$^{-1.25}$
(this is close to the findings of Eckart \& Genzel (1996)). 
The obtained critical density can be compared to the 
model density of VD2001 (Fig.~\ref{fig:model_tidal}).
For these clouds only selfgravitation and thermal pressure
of the neutral and ionized gas are taken into account.
Since our model clouds are on the edge of gravitational
instability, their central density is close to the critical
density for gravitational collapse (Fig.~\ref{fig:model_tidal}
dotted line). As long as this central density is 
higher than the critical central density with
respect to tidal shear (Fig.~\ref{fig:model_tidal} solid line),
non-rotating isothermal clouds can exist. The crossing of the
curves corresponds thus to the minimum distance where stable
clouds can exist. This critical Galactic radius is $R_{\rm
crit}\sim 3.5$~pc for this estimate. More detailed
calculations show $R_{\rm crit}\sim 2.5$~pc for the illuminated
side and $R_{\rm crit} \leq 3.5$~pc for the shadowed side.
This difference is due to the different mechanisms 
that create the cloud boundary at the illuminated and shadowed side.
At the illuminated side, the density of the ionized gas is 
given by the ionization--recombination equilibrium. 
The heated and ionized gas
flows away from the cloud and fills the space between the clouds.
In this way, a low density ionized interclump medium 
($n_{\rm e} \sim 10^{3}$~cm$^{-3}$) is built up. 
The external pressure of this interclump medium
is responsible for the outer edge of the cloud at the
shadowed side.

We have argued in the previous Sections that for a typical rotating
or magnetized cloud
$\rho_{\rm c}^{\rm rot/mag} \sim 3\times \rho_{\rm c}^{\rm BE}$
(Fig.~\ref{fig:model_tidal} dashed line). The presence of rotation
and/or magnetic fields alters the properties of the cloud, 
i.e. they are more massive and
have higher central densities. The critical Galactic radius thus
decreases to $R_{\rm crit}\sim 2$~pc. In the framework of our
model and its approximations, this is in agreement with the
observed inner edge of the CND at $R_{\rm G}\sim 1.7$~pc.
The dependence of the inner edge on the UV radiation field 
will be discussed in Sect.~\ref{sec:edge}.

\section{The influence of the stellar winds}

The stellar wind emanating from the central He{\sc i}
star cluster which is responsible for the UV radiation field
exerts a ram pressure on the illuminated side of the cloud:
\begin{equation}
P_{\rm ram} \simeq 2.5\,10^{-8}\,\frac{(\dot{M}/3\,10^{-3}\ {\rm
M}_{\odot}\,{\rm yr}^{-1}) (v/600\ {\rm
km\,s^{-1}})}{4\,\pi\,(R/2\ {\rm pc})^{2}}\ {\rm erg\,cm^{-3}}\ ,
\end{equation}
(Yusef-Zadeh \& Wardle 1993), where $\dot{M}$ is the stellar mass
loss rate due to a wind of velocity $v$. The energy density due to
the selfgravity of the cloud is given by
\begin{equation}
P_{\rm grav} \simeq 5\,10^{-8}\,(n_{\rm cl}/10^{6}\ {\rm cm}^{-3})
\frac{(M_{\rm cl}/30\ {\rm M}_{\odot})G}{(r_{\rm cl}/0.05\ {\rm
pc})} \ {\rm erg\,cm^{-3}}\ ,
\end{equation}
where $M_{\rm cl}$ and $r_{\rm cl}$ are the cloud mass and mean
radius. Ram pressure is thus able to push the ionized and heated
gas ($n_{i} \sim$ several 10$^{3}$ cm$^{-3}$) radially away from
the Galactic Centre, and might even shape the neutral 
condensations at the inner edge of the CND. It represents
an additional external force on the illuminated side of
clouds at the inner edge of the CND.

\section{The inner edge of a CND \label{sec:edge}}

If we assume a spectrum of clouds moving around a galactic centre
forming a disk-like equilibrium structure, there are four effects
that determine the physical properties of the clouds: UV
radiation, tidal shear, a radially-directed wind, and
selfgravitation. In the following considerations we will
neglect the effects of stellar winds. Nevertheless, one has to keep 
in mind that they can provide an additional external
pressure on the clouds at the inner edge of the CND.
Only those clouds with a sufficiently high central
density can resist tidal disruption. Thus, the clouds' mean
density must increase with decreasing distance to the
Galactic Centre. If they reach the Jeans mass they become
gravitationally unstable and collapse. Taking these two effects
together we obtain an efficient mechanism to create an inner edge.
At this distance, the clouds that can resist tidal disruption
become Jeans unstable, i.e. the dense cloud structure is lost. The
UV radiation plays an important r\^{o}le in determining the radius
of the clouds at each distance from the galactic centre. With an
increasing UV radiation field the cloud radius decreases, because
the ionization front in direction of the central star cluster 
is located at smaller cloud radii (Dyson 1968). 
At these radii the enhanced pressure of the ionized gas in the
ionization front is counterbalanced by the enhanced thermal
pressure due to the increasing density of the neutral gas
in the isothermal cloud. Thus the
cloud mass decreases with increasing UV radiation field
and the cloud is less susceptible to gravitational
collapse.

We can estimate this effect quantitatively in the following way. 
In the case of a cloud with uniform density the
maximum density needed to stabilize a cloud against tidal forces is
\begin{equation}
\rho_{\rm crit}^{\rm tidal}=\frac{3}{2\pi}\frac{M(R)}{R^{3}},
\label{eq:a31}
\end{equation}
where $M(R)$ is the enclosed mass up to the radius $R$ from the
Galactic Centre. The criterion for Jeans instability is given by
\begin{equation}
\rho_{\rm crit}^{\rm Jeans}=\frac{\pi{\cal R}T}{G \mu r_{\rm
cl}^{2}}\ ,
\end{equation} \label{eq:a32}
where $T$ is the temperature of the cloud, ${\cal R}$ is the gas
constant, $G$ is the gravitational constant, and $\mu$ is the
molecular weight.

The cloud radius due to the balance of ionization and
recombination is given by Dyson (1968)
\begin{equation}
r_{\rm cl}=\xi^{2} J_{0} n_{\rm i}^{-2}\ ,
\end{equation} \label{eq:a34}
where $J_{0}$ is the number of incident UV photons per cm$^{2}$
and s, $n_{\rm i}$ is the number density of the ionized gas in the
ionization front, and $\xi=4.87\ 10^{6}$ cm$^{-\frac{3}{2}}$ s$^{\frac{1}{2}}$.
With the jump condition across the ionization front $\rho_{\rm cl}
{\rm e}^{-u} c_{\rm s}^{2} = 2 \rho_{\rm i} c_{\rm i}^{2}$, where
$\rho_{\rm i}$ and $c_{\rm i}$ are the density and the sound
velocity of the external ionized gas and $u(x)$ is a parameter of the
Lane-Emden equation, one can write
\begin{equation}
r_{\rm cl}=4 \xi^{2}m_{p}^{2}J_{0}\rho_{\rm cl}^{-2}{\rm
e}^{2u}(\frac{c_{\rm i}}{c_{\rm s}})^{4}\ .
\label{eq:a35}
\end{equation}
Furthermore, one can approximate the function ${\rm e}^{-u(x)}\sim
3x^{-2}$ for $2 < x < 8$, where 
$x=r_{\rm cl}/c_{\rm s}\sqrt{4\pi G \rho_{\rm c}}$. This leads to
\begin{equation}
r_{\rm cl}=3.645\ 10^{15} J_{0}^{-\frac{1}{3}}c_{\rm
i}^{-\frac{4}{3}} c_{\rm s}^{\frac{8}{3}}\ .
\label{eq:a36}
\end{equation}
Thus, the cloud radius depends only on the number of incident
UV photons, the sound speed of the ionized gas, and the sound speed
of the neutral gas. 
If we assume that the temperature is determined by the radiation
field $c_{\rm s} \propto T^{\frac{1}{2}} \propto
J_{0}^{\frac{1}{8}}$, the cloud radius does not change
with the cloud's distance to the Galactic Centre.
This means that the cloud's radius is constant and does not depend 
on its central density. Consequently, the critical Jeans
density $\rho_{\rm crit}^{\rm Jeans}$ is proportional to the
gas temperature of the neutral gas.

\begin{figure}
    \resizebox{\hsize}{!}{\includegraphics{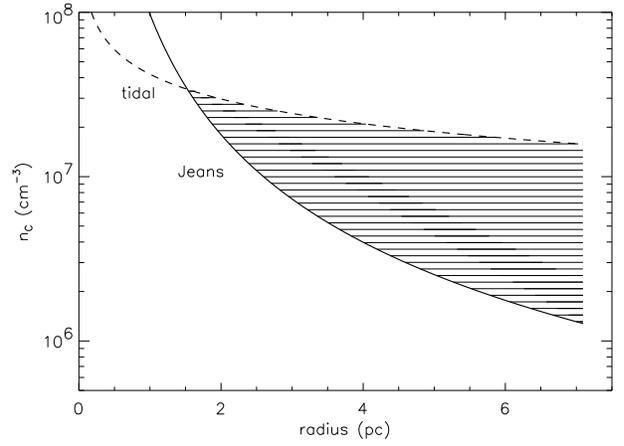}}
      \caption{Central density of the heavy clouds versus the
      distance to the Galactic Centre. Dashed line: maximum central
      density above which gravitational collapse occurs. Solid
      line: minimum density in order to resist tidal shear. Dashed
      surface: range of densities where clouds are gravitationally
      and tidally stable. }
\label{fig:grenze}
\end{figure}

Fig.~\ref{fig:grenze} shows the critical densities with respect
to tidal shear and gravitational collapse. These graphs represent
an approximation of the density variations calculated from
the detailed model of VD2001 (Fig.~\ref{fig:model_tidal}). 
The increase or decrease of the neutral gas temperature by a 
factor 2 results in a variation of the location of the 
inner edge of $\pm$1~pc.
In Fig.~\ref{fig:grenze} the dashed line shows the maximum 
central density above which gravitational collapse occurs ($\rho_{\rm
crit}^{\rm Jeans}$ using Eq.~\ref{eq:a36}), the solid line shows
the minimum density in order to resist tidal shear ($\rho_{\rm
crit}^{\rm tidal}$). The dashed surface shows the range of
densities where clouds are gravitationally and tidally stable.
The inner edge of the CND arises thus naturally as a selection
effect due to external conditions of the environment on the cloud
spectrum.

If there are non-negligible non-thermal pressure components, 
we have to add these components to the sound speed of the neutral
and ionized gas. For a non-thermal linewidth $c > 1$~km\,s$^{-1}$,
the critical Jeans density is always lower than the critical 
tidal density, i.e. no stable clouds can exist. Therefore, we suggest that
only rotating clouds, which have a higher central density can survive.

\subsection{A possible scenario for star formation}

In the previous Sections we have investigated an isolated 
clumpy gas disk. 
However, the CND appears to interact with the surrounding gas.
Coil \& Ho (1999, 2000) and Zylka et al. (1999)
conclude on the basis of the gas distribution and kinematics
in the inner 20~pc of the Galactic Centre that there are
connections between the CND and the neighbouring GMCs.
They claim that there are several streamers that fall into the
Galactic Centre.

{From the theoretical point of view} 
Sanders (1998) pointed out the possibility that the CND can be
understood in terms of tidal capture and disruption of gas clouds
falling into the Galactic Centre region. The infalling gas forms a
tidally stretched filament intersecting itself. After several
rotation periods the gas forms a stable ring structure which can
be maintained for more than 10$^{6}$ yr. He showed that the
central star cluster can be created within the first few passages
of the cloud when the long filament intersects itself at a large
angle.

We will now discuss what happens when a cloud complex falls
from a distance greater than 10\,pc into the Galactic Centre.

In VD2001 we have shown that the CND has a lifetime of $\sim
10^{7}$~yr. It is thus possible that an external cloud is falling
into the Galactic Centre within this period. We propose a new
scenario in which a whole cloud complex is falling into the
Galactic Centre where a clumpy disk structure already exists. When
the cloud hits the CND, frequent partially inelastic cloud--cloud
collisions will create a whole transient spectrum of clouds with
different masses. Those clouds which have masses above the Jeans
limit will collapse and eventually form stars. The massive stars
are thus formed within a very short time during the collision of
the cloud complex and the disk. Partially inelastic collisions
result in a decrease of the cloud velocities. Low velocity clouds
($v < v_{\rm rot} \sim 120$ km\,s$^{-1}$) form stars and approach
the Galactic Centre at the same time. Thus, if an external
gas cloud collides with a pre-existing CND, we expect that a part of
the gas, which loses angular momentum due to cloud--cloud collisions
spirals inwards with a velocity of $\sim$100~km\,s$^{-1}$.

If we assume an initial He\,{\sc i} star mass of $M_{0}=30$
M$_{\odot}$ with an initial velocity of $V_{0}=100\,$km\,s$^{-1}$,
falling into an already existing old stellar cluster of density
$\rho_{\rm stellar}=4\,10^{6}\ {\rm M}_{\odot}{\rm pc}^{-3}$ with
stellar masses around 1~M$_{\odot}$, collective relaxation
dominates over the particle-particle relaxation
(Saslaw 1985). This leads to a relaxation time of\\
\begin{equation}
\tau_{\rm R, coll}=7\, 10^{6}\times \Big (\frac{V_{0}}{100\ {\rm
km\,s}^{-1}}\Big )^{3} \Big (\frac{M_{0}} {30\ {\rm
M}_{\odot}}\Big )^{-1} \Big (\frac{\rho_{\rm stellar}} {4\,10^{6}\
\frac{\rm M_{\odot}}{\rm pc^{3}}}\Big )^{-1} {\rm yr}\ .
\label{eq:a37}
\end{equation}
A single massive star, whose velocity is approximately the
Keplerian velocity of gas moving on circular orbits around the 
Galactic Centre, thus loses the information about its initial
position and velocity within $\sim$7\,10$^{6}$ yr. 

Gerhard (2000) estimated that a star cluster of $\sim
10^{5}$~M$_{\odot}$ which is formed at a Galactic radius $R_{\rm
G}$=10~pc needs several Myr to spiral into the Galactic Centre.
This timescale is comparable to our collective relaxation
timescale.

The star cluster relaxation time is thus comparable to the
lifetime of the He{\sc i} stars. These massive stars are formed at
the same time and at the same distance to the Galactic Centre
forming a star cluster after a few million years. The observed
streaming motions of the He{\sc i} stars in the Galactic Centre
(Genzel et al. 1996) shows that the star cluster is not completely
relaxed. Therefore, it is possible that the central He{\sc i} star
cluster has been built during the collision of an infalling cloud
complex with an already existing CND.

\section{Conclusion}

We investigated the influence of rotation and magnetic fields
on the physical properties of the gas clouds located in the 
Circumnuclear Disk in the Galactic Centre. 
Rotating selfgravitating isothermal clouds of a given
temperature $T$ embedded in a tenuous medium giving rise to an
external pressure $P_{\rm ext}$ are more massive and have larger
densities than non-rotating clouds of same $T$ and $P_{\rm ext}$.
Stable rotating clouds have an angular momentum in the range
$3\,10^{-2} \leq J/M \leq 5\,10^{-2}$~pc\,km\,s$^{-1}$. This
represents $\sim$20\%--100\% of the maximum specific angular momentum
that a cloud can acquire during its formation. These clouds are
stable against tidal shear for Galactic Radii $R_{\rm G} \geq
2$~pc. Magnetized selfgravitating isothermal clouds with magnetic
field strengths of several mG have a critical mass with respect to
gravitational collapse of $\sim$3 times the corresponding
Bonner--Ebert critical mass of a selfgravitating isothermal sphere.

We suggest a mechanism for the formation of an inner edge in
circumnuclear disks. The external UV radiation field determines
the diameter of the clouds. The density of the clouds must
increase with decreasing galactic radius because of the tidal
shear due to the gravitational potential in the Galactic Centre.
At a critical radius clouds that are stable against tidal
shear become gravitational unstable and collapse. On the basis
of our model, selfgravitating non-magnetized isothermal clouds
should rotate in order to resist tidal shear at a distance of
2~pc from the Galactic Centre, whereas magnetized clouds must
rotate in order to resist tidal shear. For the
CND in the Galactic Centre the critical radius corresponds
approximately to the observed inner edge if cloud rotation is
included. We suggest that the central He{\sc i} star cluster
has been formed when an external cloud collided with a pre-existing
CND.

\begin{acknowledgements}
We would like to thank P. Ho for helping us to improve this
article significantly.
\end{acknowledgements}

\end{document}